\documentclass[11pt, a4paper]{article}
\usepackage{latexsym}
\usepackage{indentfirst}
\usepackage{graphicx}
\usepackage{amsmath}
\usepackage{amssymb}

\usepackage{appendix}
\usepackage{amstext}    
\usepackage{array} 
\usepackage{booktabs} 
\usepackage{caption}
\usepackage{subcaption}
\usepackage{epstopdf}

\begin{document}
\bibliographystyle{alpha}
\title{On the impact of explicit or semi-implicit integration methods over the stability of real-time numerical simulations}
\author{Teodor Cioac\u a$^1$, Horea C\u ar\u amizaru$^2$ \\ $^{1,2}$"Politehnica" University Bucharest \\
$^1$ t.cioaca@gmail.com}

\date{}
\maketitle

\begin{abstract}
Physics-based animation of soft or rigid bodies for real-time applications often suffers from numerical instabilities. 
We analyse 
one of the most common sources of unwanted behaviour: the numerical integration strategy. 
To assess the impact of popular integration
methods, we consider a scenario where soft and hard constraints are added to a custom designed deformable linear object. 
Since the goal for this class
of simulation methods is to attain interactive frame-rates, we present the drawbacks of using explicit integration methods over inherently
stable, implicit integrators. 
To help numerical solver designers better understand the impact of an integrator on a certain simulated world, 
we have conceived a method
of benchmarking the efficiency of an integrator with respect to its speed, stability and symplecticity.

\end{abstract}

\section{Introduction}
Intricate physical and mathematical models for simulating soft or rigid bodies are the centre of intensive research efforts. Computer animation, laparoscopic haptic surgery simulations,
graphical special effects or robotic manipulation are just a few of the fields where such models play a key role. The focus of our research is to address the issue of carrying out stable, physics-based simulations at interactive update rates. For this purpose, we have conceived an elementary framework
for building deformable objects with soft or stiff constraints. 
Using this framework, we can test the efficiency and impact of several, well-known
numerical integrators. 
To address our goal, we look find an equilibrium between an integrator's computational overhead, its precision, symplecticity and, most importantly, its inherent stability.

Following the aforementioned goals, we present a short literature survey on the numerical simulation of soft or semi-rigid
bodies in section \ref{sec:related_work_CC}. The mathematical
apparatus for creating and simulating constrained objects is explained in section \ref{sec:constrained_animation_CC},
and our own, deformable linear object model case study is
presented in section \ref{sec:dlo_CC}. We present a general update strategy that supports any type of explicit
or semi-implicit integration method in section \ref{sec:simulation_CC}. Finally,
the efficiency of several popular integrators is analysed and discussed in section \ref{sec:results_CC},
and we conclude this research in section \ref{sec:conclusion_CC}.
 
\section{Related work}
\label{sec:related_work_CC}
Stability as a central attribute to Physics-based object simulation within the Computer Graphics world was addressed by directly consider inherently stable integrators.  The most popular method employed is the first order, implicit Euler scheme. Servin et al. \cite{Servin151856} treated infinitely stiff springs as kinematic constraints and developed a system capable of simulating elastic behaviour at the cost of solving a large sparse linear system for each iteration.  The hybrid method of Schroeder et al. \cite{journals/cgf/SchroederKZF11} uses explicit updates for the elastic forces and implicit strategies for other components. This complex idea tries to reduce the effect of suppressing material vibrations due to the use of pure implicit integrators. By alternating between implicit, semi-implicit and explicit methods, Volino et al. \cite{Volino2005} demonstrated how different cloth materials can be modeled to overcome the inabilities of one integrator to support a certain material property. Some object 
models allow using for computing an analytical force approximation for an implicit solver as demonstrated by Mesit et al. \cite{journals/jcp/MesitGC07}. Their method of simulating gas filled soft bodies can thus easily support any integrator, but it relies on the topological structure for all analytical derivations.
Finally, structured mass-spring models using pure implicit integrators were popularized by Baraf and Witkin \cite{Baraff:1998:LSC:280814.280821} in their seminal paper on stable cloth simulation and by Desbrun et al. \cite{Desbrun:1999:IAS:351631.351638} where inverse dynamics were used to tackle outstretching artefacts. For a discussion of the performance of some popular implicit solvers we direct the user to  the paper of Hauth et al. \cite{journals/cgf/HauthE01}.

Explicit methods are more popular due to their reduced complexity. Finite element simulations using explicit integrators were performed by Fierz et al. \cite{eth_biwi_00802}. To stabilize the numerical process, the authors proposed modifying the stiffness matrices of ill-conditioned tetrahedral elements. This allowed their simulation to use higher time steps in spite of the explicit updates. Although less computationally demanding and stable than their implicit counterparts, explicit integrators can share a fair amount of stability, energy conserving capabilities and time reversibility. In this respect, Tsai et al. \cite{TSAI2004} present a comprehensive survey on symplectic integrators used in molecular dynamics.

In a different class of their own, position-based dynamics (see Bender et al. \cite{BMOT2013} for a survey) offer a workaround for any force or impulse based system, avoiding the intricacies of using integrators. This family of methods is, however, inaccurate for scenarios where velocities and forces  need to be measured, hence we mention it as an alternative for special effects applications.

For a comprehensive list of simulation methods involving deformable objects, we invite the reader to consult the work of Nealen et al. \cite{CGF:CGF1000} or that of Jimenez \cite{Jimenez2012}.

\section{Constrained object animation}
\label{sec:constrained_animation_CC}
To better understand the elements involved in most physics-based animation scenarios, we will briefly introduce a simple constraint enforcing system. Supporting this kind of simulation mechanism requires a discrete geometrical sampling of the initial shape of the object. For exemplification purposes, we consider a deformable linear object, embedded in a 3D space, whose cross-section is considerably smaller than its length. Its elastic
properties that determine how its discrete structure changes are implemented by adding geometric constraints enforced by potentials (see the work of Teschner et al. \cite{Teschner2004} for a more detailed and generalized application). These potentials provide a direct measure of how the structure of points differs from its initial 
configuration.

Generally, if $\mathbf{p}_1, \ldots, \mathbf{p}_N$ are the vertices of a constrained group, then a constraint function is defined as follows:
\begin{equation}
\label{eq:constraint_function_CC}
C(\mathbf{p}_1, \ldots, \mathbf{p}_N): \mathbb{R}^{3N} \to \mathbb{R}.
\end{equation}
This function incorporates additional information relating the current vertex configuration of the group to the initial 
geometrical image through specific scalar measures of length, area, angle, volume, etc. For example, a length based constraint
incorporates the initial or rest lengths of directly connected vertices as a reference for the measure of deviation.  Generally, an energy function produces only positive amounts and can be written as:

\begin{equation}
\label{eq:constraint_potential_CC}
E(\mathbf{p}_1, \ldots, \mathbf{p}_N) = \frac{1}{2} C(\mathbf{p}_1, \ldots, \mathbf{p}_N)^2 .
\end{equation}
For a $\{\mathbf{p}_1, \ldots, \mathbf{p}_N\}$ configuration, we now consider the restriction of the energy function at a node $\mathbf{p}_i$. This function is written as:
\begin{equation}
E_{\mathbf{p}_i}(\mathbf{x}) : \mathbf{x} \in \mathbb{R}^3 \to \mathbb{R}_{+} .
\end{equation} 

Since the gradient vector coincides with the direction of maximum potential increase, it is natural to consider a penalty function that points in the opposite direction:
\begin{equation}
\label{eq:penalty_function_CC}
F_{\mathbf{p}_i}(\mathbf{x}) = - \nabla E_{\mathbf{p}_i}(\mathbf{x}).
\end{equation}
This last equation can be written in an equivalent form:
\begin{equation}
F_{\mathbf{p}_i}(\mathbf{x}) = -C_{\mathbf{p}_i}(\mathbf{x}) \nabla{C_{\mathbf{p}_i}(\mathbf{x})}.
\end{equation}

The simplest example of such a behaviour is the case of a linear spring connecting two points $\mathbf{p}_i$ and $\mathbf{p}_j$. The natural constraint function  is defined as:
\begin{equation}
C(\mathbf{p}_i,\mathbf{p}_j) = \|\mathbf{p}_i - \mathbf{p}_j \| - L_0 ,
\end{equation}
where $L_0$ denotes the spring's rest length. From this constraint function, a corresponding elastic deformation potential energy can be expressed as:
\begin{equation}
\label{eq:elastic_energy_CC}
E(\mathbf{p}_i,\mathbf{p}_j) = \frac{1}{2} \left( {\|\mathbf{p}_i - \mathbf{p}_j\|| - L_0} \right)^ 2.
\end{equation}
We define the following operator:
\begin{equation}
\frac{\partial}{\partial \mathbf{p}_i} E = \nabla E_{\mathbf{p}_i} ,
\end{equation}
also known in some works as the variational or functional derivative operator.

An elastic spring force at node $\mathbf{p}_i$ can be derived using equation \ref{eq:elastic_energy_CC}:
$$ F_{\mathbf{p}_i}^{spring}(\mathbf{p}_i, \mathbf{p}_j) = -\frac{\partial}{\partial \mathbf{p}_i} \left( \frac{1}{2} \left( \|\mathbf{p}_i - \mathbf{p}_j\|| - L_0 \right)^ 2 \right) .$$

After applying several derivative computation rules (chain rule and the derivative of a product of two functions), we find the general elastic force expression:
\begin{equation}
\label{eq:elastic_force_CC}
F_{\mathbf{p}_i}^{spring}(\mathbf{p}_i,\mathbf{p}_j) =  - \mathrm{K_l}\left( \|\mathbf{p}_i - \mathbf{p}_j\|| - L_0 \right) \frac{(\mathbf{p}_i - \mathbf{p}_j)}{\| \mathbf{p}_i - \mathbf{p}_j\|},
\end{equation}
where $\mathrm{K_l}$ is the linear spring's stiffness coefficient. 


A deformable object defined as a closed surface can be discretized by dividing its interior volume into tetrahedral cells. Apart from linear springs, it is desirable to enforce local constraints aimed at preserving the volumes under deformation. Such forces are easy to introduce by deriving them from a volume preserving constraint function involving tetrahedral cells:
\begin{equation}
\label{eq:voume_constraint_CC}
\begin{split}
& C_V(\mathbf{p}_i, \mathbf{p}_j, \mathbf{p}_k, \mathbf{p}_l) = \frac{\frac{1}{6}\left\langle \mathbf{p}_j - \mathbf{p}_i, (\mathbf{p}_k - \mathbf{p}_i) \times (\mathbf{p}_l - \mathbf{p}_i)\right\rangle - V_0}{V_0},
\end{split}
\end{equation}
where $V_0$ is the volume of the tetrahedron in its rest configuration.
Subsequently, the volume preserving potential is:
\begin{equation}
\label{eq:volume_potential_CC}
E_V(\mathbf{p}_i, \mathbf{p}_j, \mathbf{p}_k, \mathbf{p}_l) = \frac{1}{2} C_v(\mathbf{p}_i, \mathbf{p}_j,\mathbf{p}_k,\mathbf{p}_l)^2 .
\end{equation}
Using the same method as for linear springs, we compute the force at the $\mathbf{p}_i$ vertex by using the $\frac{\partial}{\partial \mathbf{p}_i}$ operator:
\begin{equation}
\label{eq:volume_force_CC}
F_{V_{\mathbf{p}_i}}(\mathbf{p}_i, \mathbf{p}_j, \mathbf{p}_k, \mathbf{p}_l) = - \frac{\partial}{\partial \mathbf{p}_i} E_V(\mathbf{p}_i, \mathbf{p}_j, \mathbf{p}_k, \mathbf{p}_l) .
\end{equation}
After conveniently arranging the results from the derivation of equation  (\ref{eq:volume_force_CC}), we can write down the expanded formula of this force:
\begin{equation}
\label{eq:volume_preserving_force_CC}
\begin{split}
 F_{V_{\mathbf{p}_i}}(\mathbf{p}_i, \mathbf{p}_j, \mathbf{p}_k, \mathbf{p}_l) = & \frac{K_V}{6V_0^2}\left[\frac{1}{6}(\mathbf{p}_j - \mathbf{p}_i) \cdot (\mathbf{p}_k - \mathbf{p}_i) \times (\mathbf{p}_l - \mathbf{p}_i) - V_0\right] \cdot \\
& \cdot (\mathbf{p}_j - \mathbf{p}_l) \times (\mathbf{p}_k - \mathbf{p}_l),
\end{split}
\end{equation}
where $K_V$ is an added-in stiffness coefficient. 
\begin{figure}[h!]
\centering
\includegraphics[width=0.45\textwidth]{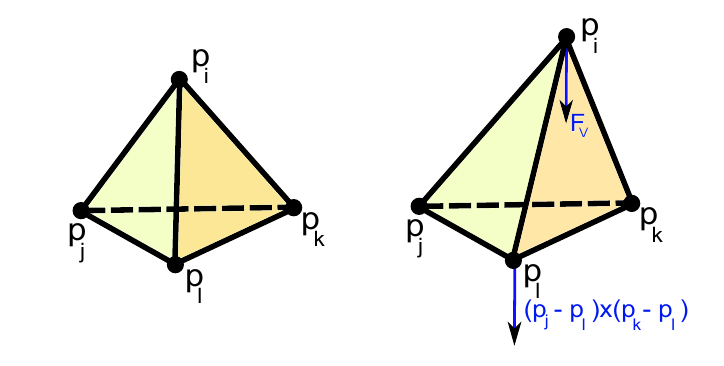}
\caption{Volumetric force on a tetrahedral cell.}
\label{fig:dlo_volumetric_force_CC}
\end{figure}
Examining figure \ref{fig:dlo_volumetric_force_CC}, we can see how the volumetric force acts to prevent volume changes. For example, if point $\mathbf{p}_i$ is shifted and the volume 
increases as a consequence, the direction of the volumetric force is given by the cross product vector $(\mathbf{p}_j - \mathbf{p}_l) \times (\mathbf{p}_k - \mathbf{p}_l)$.
Inconsistent or "flipped" cell configurations, as well as degenerate tetrahedra can and are likely to be encountered during a simulation. Compressible bodies prevent themselves from being completely flattened by acting 
as non-linear spring elements. When their state is close to a collapse, the elastic forces should increase asymptotically towards
infinity. Since the volumetric force component acts like a vertex to face spring in our system, we add a non-linear spring component, as suggested in \cite{Cooper97}. Hence, if $V$ is the current volume of a $(\mathbf{p}_i,\mathbf{p}_j,\mathbf{p}_k,\mathbf{p}_l)$ tetrahedron, then we can write the expression of the improved volumetric force as:
\begin{equation}
\label{eq:nonlinear_volumetric_force_CC}
\begin{split}
& F_{V_{\mathbf{p}_i}}(\mathbf{p}_i, \mathbf{p}_j, \mathbf{p}_k, \mathbf{p}_l) = \\
& K_V \left[ \frac{V - V_0}{6V_0^2} - \sigma(V_0)\frac{V_0^2}{|V|} + V\right] \cdot (\mathbf{p}_j - \mathbf{p}_l) \times
(\mathbf{p}_k - \mathbf{p}_l),
\end{split}
\end{equation}
where $\sigma(x)$ is the sign function defined on the set of real values.

Another scenario that can prevent the simulation from recovering its initial rest shape in the absence of perturbing forces is the inversion phenomenon. The authors of \cite{Schmedding2008} present an improved mechanism, capable of handling the inversion 
of a tetrahedral structure for finite element simulations. This process relies on finding a rotation that best aligns the deformed cell with its undeformed counterpart and then deriving penalty forces. While we could have used a similar approach, the above modification works for cases where negative volumes are reported. The elastic penalty forces described by equation (\ref{eq:nonlinear_volumetric_force_CC}) are capable of acting against the inversion.
Additionally, other constraint-based forces can be derived (e.g. area preserving forces for the triangular faces of a tetrahedron or angle preserving forces).

\section{Deformable linear object model}
\label{sec:dlo_CC}
Using constrained mass point configuration, we can now describe the steps required to build a tetrahedral cell-based deformable object along the geometric image of a support curve:
\begin{enumerate}
\item \emph{A curve discretization:} $\{ \mathbf{r}_0, \ldots, \mathbf{r}_{N-1} \}$ where $ \mathbf{r}_i$ is a sample 3D point.
\item \emph{A set of frames:} define the orientation vectors $\mathbf{q}_k = \frac{\overrightarrow{\mathbf{r}_k\mathbf{r}_{k-1}} \times \overrightarrow{\mathbf{r}_k\mathbf{r}_{k+1}}}
{\|\overrightarrow{\mathbf{r}_k\mathbf{r}_{k-1}} \times \overrightarrow{\mathbf{r}_k\mathbf{r}_{k+1}}\|}$, and 
$\mathbf{p}_k = \frac{1}{\| \overrightarrow{\mathbf{r}_k\mathbf{r}_{k-1}} \|} \mathbf{q}_k \times \overrightarrow{\mathbf{r}_k\mathbf{r}_{k-1}}$. At each point $\mathbf{r}_k$, a coordinate frame $\{ \overrightarrow{\mathbf{r}_k
\mathbf{r}_{k+1}}, \overrightarrow{\mathbf{r}_k \mathbf{q}_k}, \overrightarrow{\mathbf{r}_k\mathbf{p}_k}\}$ is attached. 

\item \emph{Volumetric cells:} for each  pair of neighbouring vertices, $\mathbf{r}_k \mathbf{r}_{k+1}$, 
three tetrahedra are constructed: $(R_{k+1}P_kQ_kR_k)$, $(R_{k+1}Q_{k+1}Q_kP_k)$, and  $(R_{k+1}P_kP_{k+1}Q_{k+1})$ (as depicted in figure \ref{fig:dlo_tets_CC} ).
\begin{figure}[htb]
\centering
\includegraphics[width=0.45\textwidth]{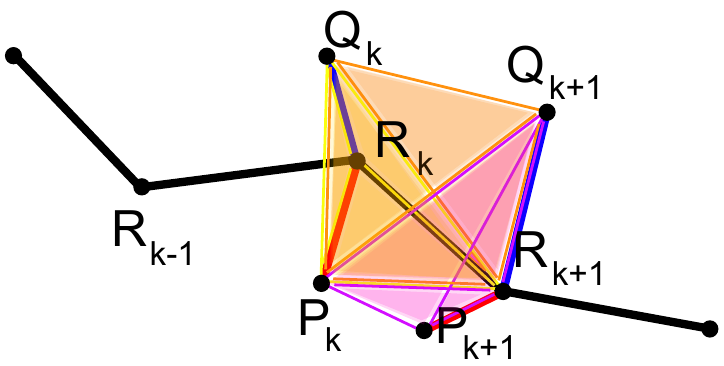}
\caption{Tetrahedral cell division of a DLO segment.}
\label{fig:dlo_tets_CC}
\end{figure}

\end{enumerate}
The tetrahedral cells  allow retrieving consistent information about local torsion and curvature changes from one link to another adjacent segment. A direct reference for how the object twists around the $R_kR_{k+1}$ line is given by the relative orientation of $\overrightarrow{R_kQ_k}$ to 
$\overrightarrow{R_{k+1}Q_{k+1}}$. Structural resistance is added by substituting the edges of the tetrahedra with linear springs.
These constraints tend to act like curvature springs since they oppose the bending of the object around the $R_kQ_k$ line.

\begin{figure}[htb]
\centering
\includegraphics[width=0.45\textwidth]{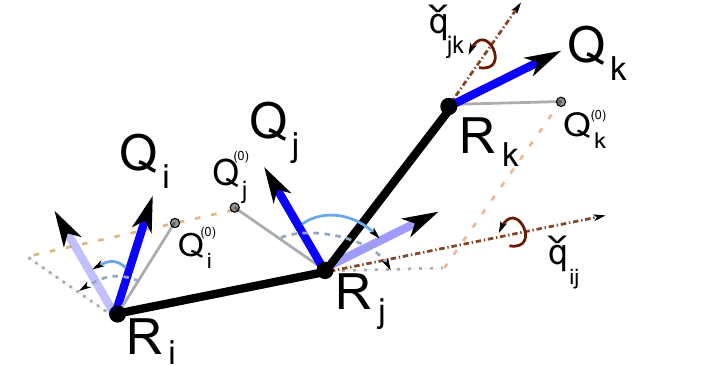}
\caption{Torsion compensation using quaternions.}
\label{fig:dlo_quat_CC}
\end{figure}

To account for plausible twisting behaviour, we introduce a quaternion based constraint system that acts like a torsional spring at
each $R_k$ node of the object. Considering there are three connected nodes, $R_i,R_j$ and $R_k$, in this order, the torsional spring tends to reposition the $Q_j$ point such that the 
resulting configuration is closer to the initial, rest configuration. Such a process requires finding a suitable axis and computing the relative orientation of the 
$\overrightarrow{R_jQ_j}$ vector with respect to the $\overrightarrow{R_iQ_i}$ and $\overrightarrow{R_kQ_k}$ vectors. We achieve this behaviour by computing a torsion compensating
quaternion:
\begin{equation}
 \tilde{\mathfrak{q}}_j  = \mathtt{SLERP}(\check{\mathfrak{q}}_{ij}, \check{\mathfrak{q}}_{jk}^{*}, \lambda) , 
\end{equation}
where $\check{\mathfrak{q}}_{ij}  = \mathtt{Quat}(\widehat{R_iR_j}, \frac{1}{2}(u_ij^{(0)} - u_{ij})),\check{\mathfrak{q}}_{jk}  = \mathtt{Quat}(\widehat{R_jR_k}, \frac{1}{2}(u_{jk}^{(0)} - u_{jk}))$, $ \lambda  = \frac{ \| \overrightarrow{R_iR_j} \| }{\| \overrightarrow{R_iR_j} \| + \| \overrightarrow{R_jR_k} \|} $. The angle $u_{ab}  = \mathtt{angle}_{\widehat{R_aR_b}}(\overrightarrow{R_aQ_a},\overrightarrow{R_bQ_b})$ represents the angle between the $\overrightarrow{R_aQ_a}$ and
$\overrightarrow{R_bQ_b}$ vectors with respect to the $\widehat{R_aR_b}$ axis , and the $^{(0)}$ superscript designates values for the initial, undeformed state of the object.
The torsion quaternion, $\check{\mathfrak{q}}_j$ can then be used to rotate the $\overrightarrow{R_jQ_j}$ vector to minimize the torsion offset. Instead of directly rotating this vector,
a force will be applied to the $Q_j$ node, thus mimicking the effects of an angular spring. Figure \ref{fig:dlo_quat_CC} depicts how the torsion quaternions are derived with respect to local geometry.

\section{Simulation update logic}
\label{sec:simulation_CC}
In general, the steps required for a complete update of the object's state can be assembled as follows:
\begin{itemize}
\item \emph{Numerical integration:} compute the current acceleration $\mathbf{a}^{(n)}$ from the current state $(\mathbf{x},\mathbf{v})^{(n)}$, and add any force contributions from
the collision solver stage. Using an explicit integration method, the new $(\mathbf{x},\mathbf{v})^{(n+1)}$ state is computed.
\item \emph{Approximate forces:} $\mathbf{f}^{(n+1)} = \frac{\mathbf{v}^{(n+1)} - \mathbf{v}^{(n)}}{\Delta t}$. These values are to be used in the collision response stage.
\item \emph{Correct positions:} applying a position based dynamics constraint projection, the linear springs are additionally relaxed. Such a process is equivalent to an iterative Gauss-Seidel solver and it requires translating the object vertices by small displacements. This step helps satisfying stiff constraints and avoids outstretching artifacts.

\item {Collision resolution:}  pairwise link collisions are detected and response forces and impulses are derived. The positions $\mathbf{x}^{(n+1)}$ and velocities $ \mathbf{v}^{(n+1)}$ are corrected, storing response forces in accumulators, $\Delta f$. These force residues are fed back to the numerical integration scheme and are included in the acceleration component the next iteration will use.
\end{itemize}
These steps must then be performed in this order, leading to a plausible update effect at interactive frame-rates.
\section{Results}
\label{sec:results_CC}
\begin{figure}[h]
        \centering
        \begin{subfigure}[b]{0.5\textwidth}
                \centering
                \includegraphics[width=\textwidth]{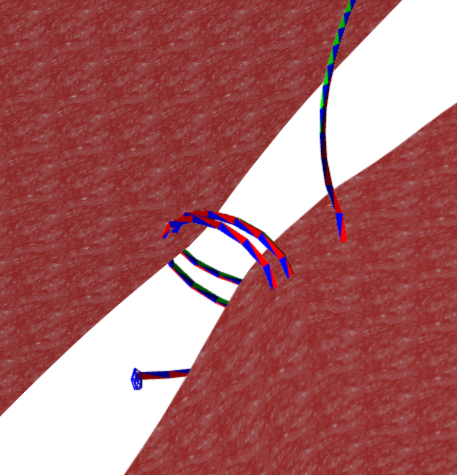}
                \caption{Driving a wire through a severed tissue layer }
                \label{fig:spiral_CC}
        \end{subfigure}%
        ~ 
        \begin{subfigure}[b]{0.5\textwidth}
                \centering
                \includegraphics[width=\textwidth]{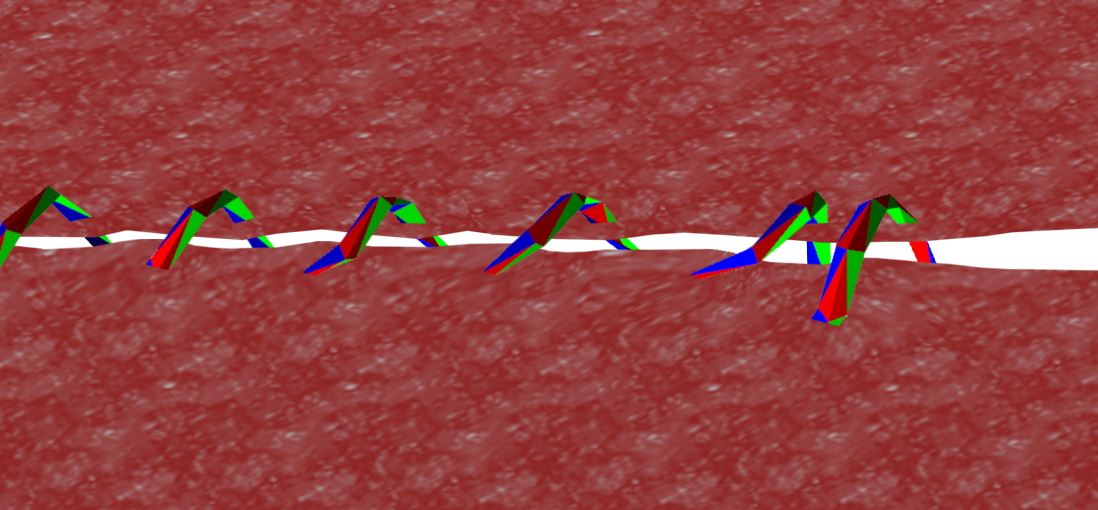}
                \caption{Tightening the suturing wire}
                \label{fig:suture_CC}
        \end{subfigure}
        \caption{Suturing simulation}\label{fig:suture}
\end{figure}

We have used our model to simulate a simple laparoscopic suturing task. Given a physical substrate, a thread was driven through a tissue-like structure (see figure \ref{fig:suture}).
The wire model we used is ideal for testing the behaviour of an integrator where both stiff and soft constraints drive the simulation. A simple methodology was employed to compute
a score sheet for each integration scheme. We tracked the speed and stability of several methods while varying the time-step. The results were quantized by considering the explicit Euler method as a reference. The order of accuracy is not particularly important as it is orthogonal to our stability goals. Due to their energy-conserving properties, symplectic integrators are
favoured when competing with other methods that achieve similar scores.

\begin{table}[h] 
\centering 
\begin{tabular}{l c c c c c } 
\toprule 
& \multicolumn{5}{c}{Criteria} \\ 
\cmidrule(l){2-5} 
Method & Max $\Delta t$ &  Time & Symplectic & Order & Score\\ 
\midrule 
Explicit Euler & 0.0098s & 8ms & NO & $O(h)$ & 1.225 \\ 
Symplectic Euler & 0.0294s & 8ms & YES & $O(h)$ & 3.675\\ 
Midpoint & 0.0153s & 8ms & NO & $O(h^2)$ & 1.9125 \\ 
Half-Step & 0.0168s & 10ms & NO & $O(h^2)$ & 1.68\\ 
Verlet & 0.023s & 10ms & YES & $O(h^2)$ & 2.3 \\ 
Forest-Ruth & 0.021s & 12ms & YES & $O(h^3)$ & 1.75 \\ 
Symplectic Midpoint & 0.028s &  8ms & YES & $O(h^2)$ & 3.5 \\ 
Runge-Kutta 4 & 0.027s & 14ms & NO & $ O(h^4)$ & 1.928 \\ 
Moified Half-Step & 0.0216s & 10ms & NO & $O(h^2)$ & 2.16 \\ 
\midrule 
\midrule 
\end{tabular}
\caption{Integrator benchmark results} 
\label{tab:benchmark_CC} 
\end{table}
Analyzing table \ref{tab:benchmark_CC}, the symplectic Euler method (discussed by Cromer \cite{cromer:455}) is the most stable for the total iteration time it needs (the score is obtained as the ratio between the maximum time step and the time needed to execute one simulation iteration). We modified the Midpoint method to achieve symplecticity, obtaining the second highest score.
Given the fact that the Midpoint family of methods is accurate to the second order for the position terms (while being a first order method for the velocity terms), we recommend it for applications where accuracy is of some importance. The Verlet method ( \cite{Verlet67} ), popular for molecular dynamics simulations, also benefits from its relatively high stability, symplecticity and second order accuracy. A third order method, the Forest-Ruth integration scheme \cite{Forest:1990:FSI:84986.84996}, represents the best alternative for applications where accuracy is a key element. The Runge-Kutta fourth order method, although supporting relatively high time steps, is the most time consuming and probably not a good choice for real time applications. 

Although symplectic integrators excel in scenarios where their energy preserving features are central (e.g. where only conservative forces are involved), we have found this family of integrators to outperform their explicit integrator counterparts. As a last remark, we have also modified the Half Step method (\cite{Hairer:1993:SOD:153158}) to support a semi-implicit update for the middle estimation $(\mathbf{x}_{n+\frac{1}{2}}, \mathbf{v}_{n+\frac{1}{2}})$. This modification significantly increases the method's stability, as seen in table \ref{tab:benchmark_CC}.

\begin{figure}[h]
        \centering
        \begin{subfigure}[b]{0.5\textwidth}
                \centering
                \includegraphics[width=\textwidth]{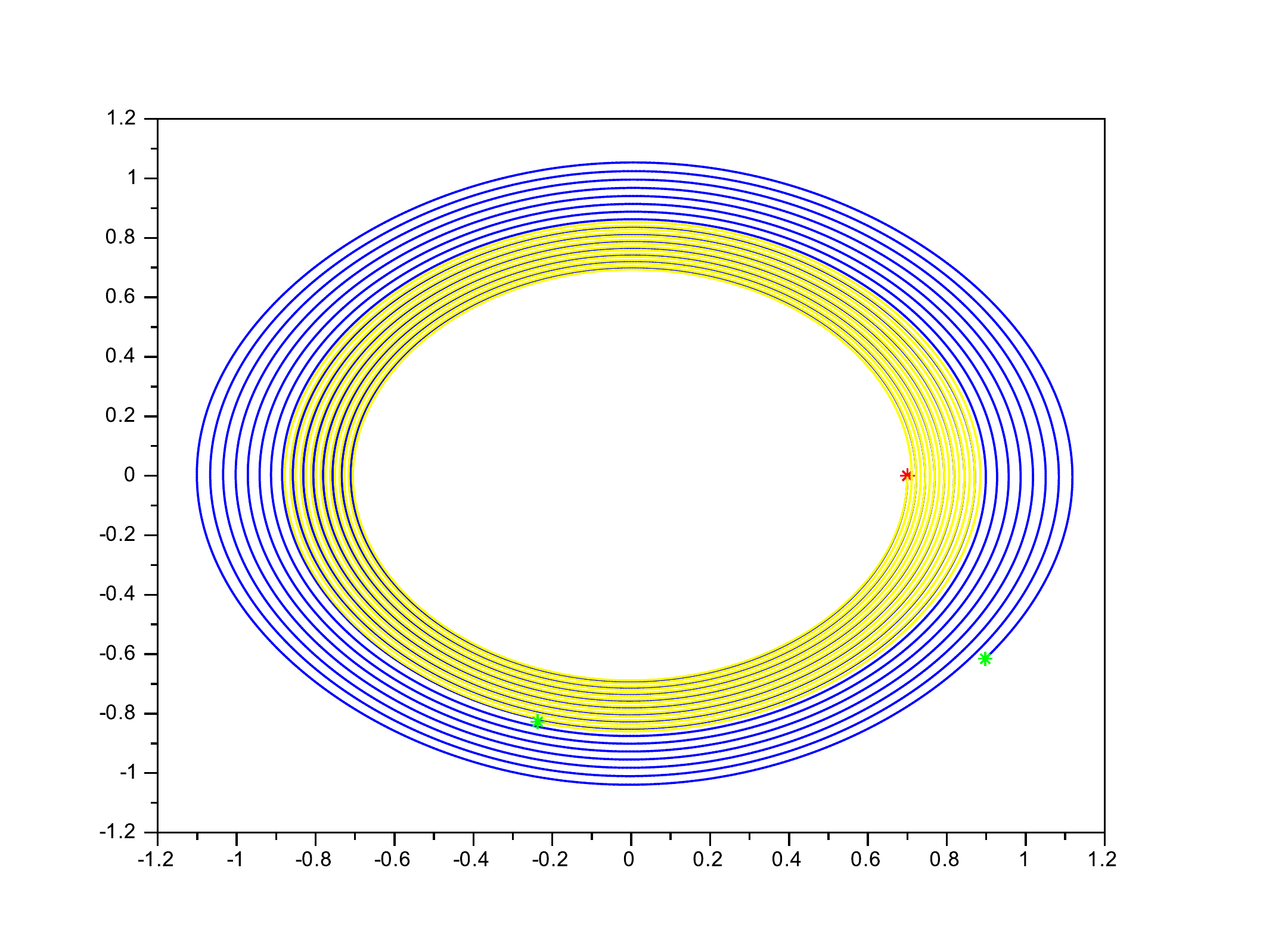}
                \caption{Explicit Euler (blue) and Midpoint (yellow), $\Delta t = 0.03$ }
                \label{fig:ee_mp_CC}
        \end{subfigure}%
        ~ 
        \begin{subfigure}[b]{0.5\textwidth}
                \centering
                \includegraphics[width=\textwidth]{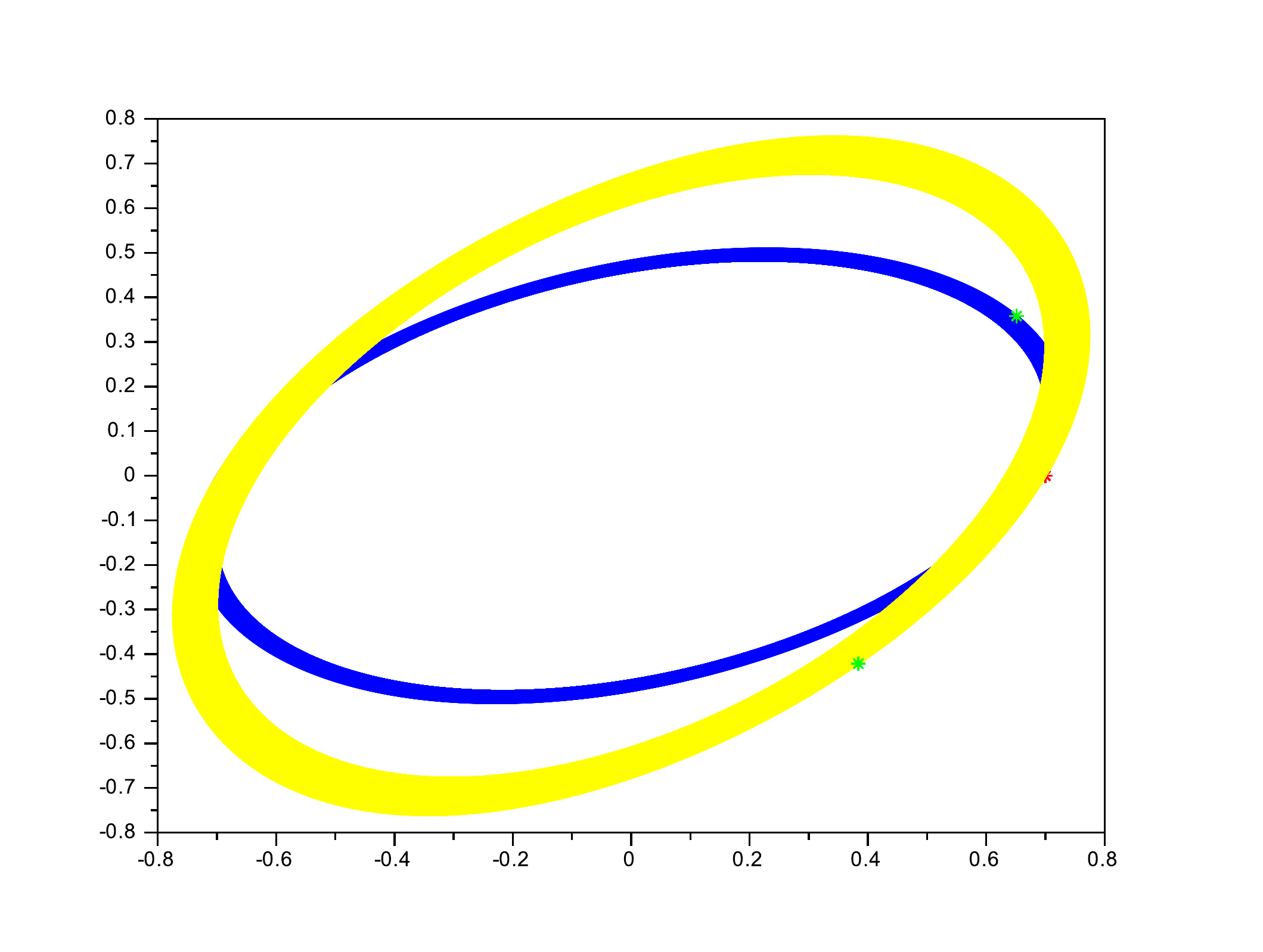}
                \caption{Symplectic Midpoint (blue) and Symplectic Euler (yellow), $\Delta t = 0.9$ }
                \label{fig:sm_se_CC}
        \end{subfigure}
        \caption{First order integrators (phase space diagrams)}\label{fig:first_order_CC}
\end{figure}

\begin{figure}[h]
        \centering
        \begin{subfigure}[b]{0.5\textwidth}
                \centering
                \includegraphics[width=\textwidth]{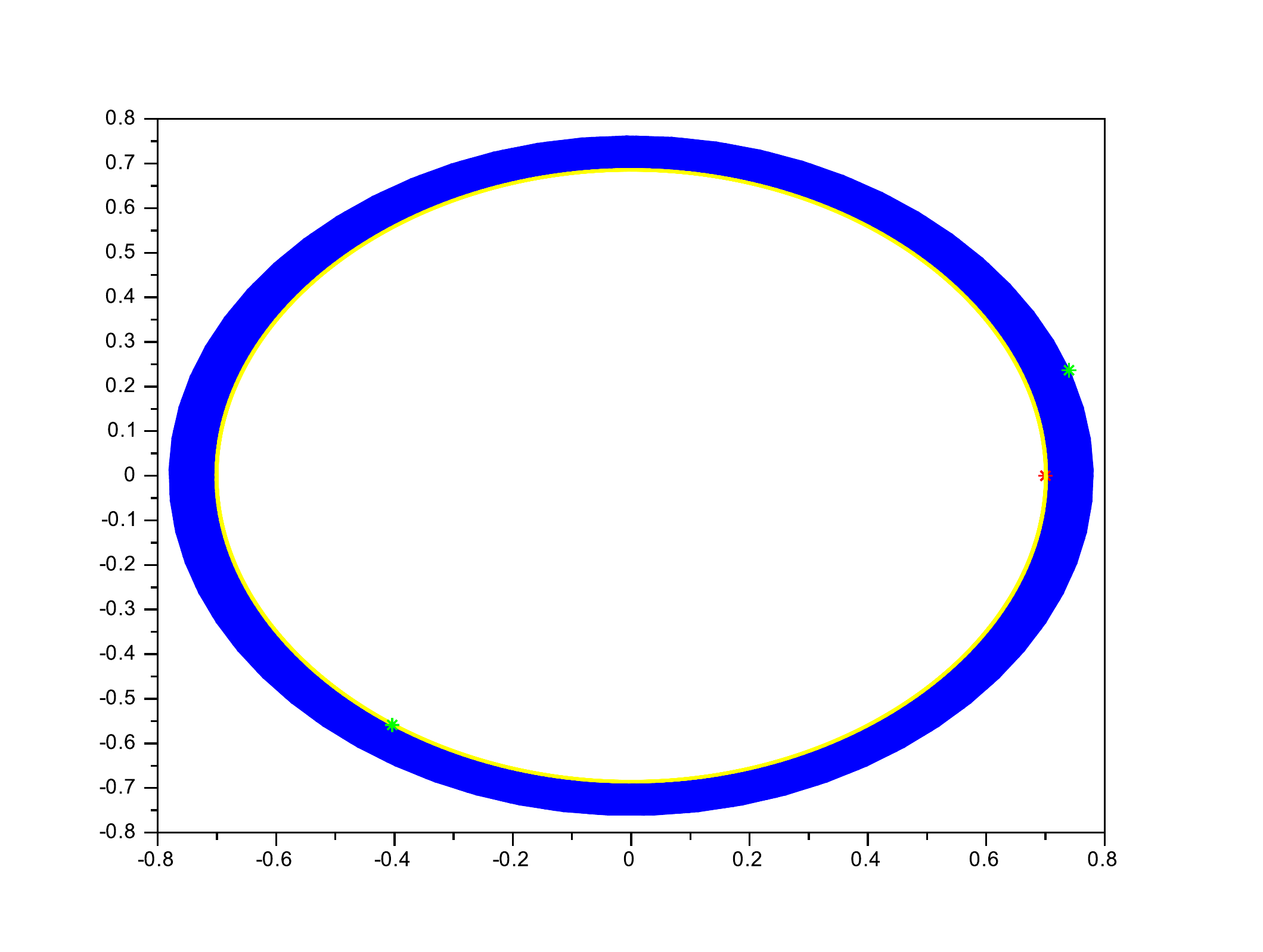}
                \caption{Half Step (blue) and Modified Half Step (yellow), $\Delta t = 0.3$ }
                \label{fig:hs_mhs_CC}
        \end{subfigure}%
        ~ 
        \begin{subfigure}[b]{0.5\textwidth}
                \centering
                \includegraphics[width=\textwidth]{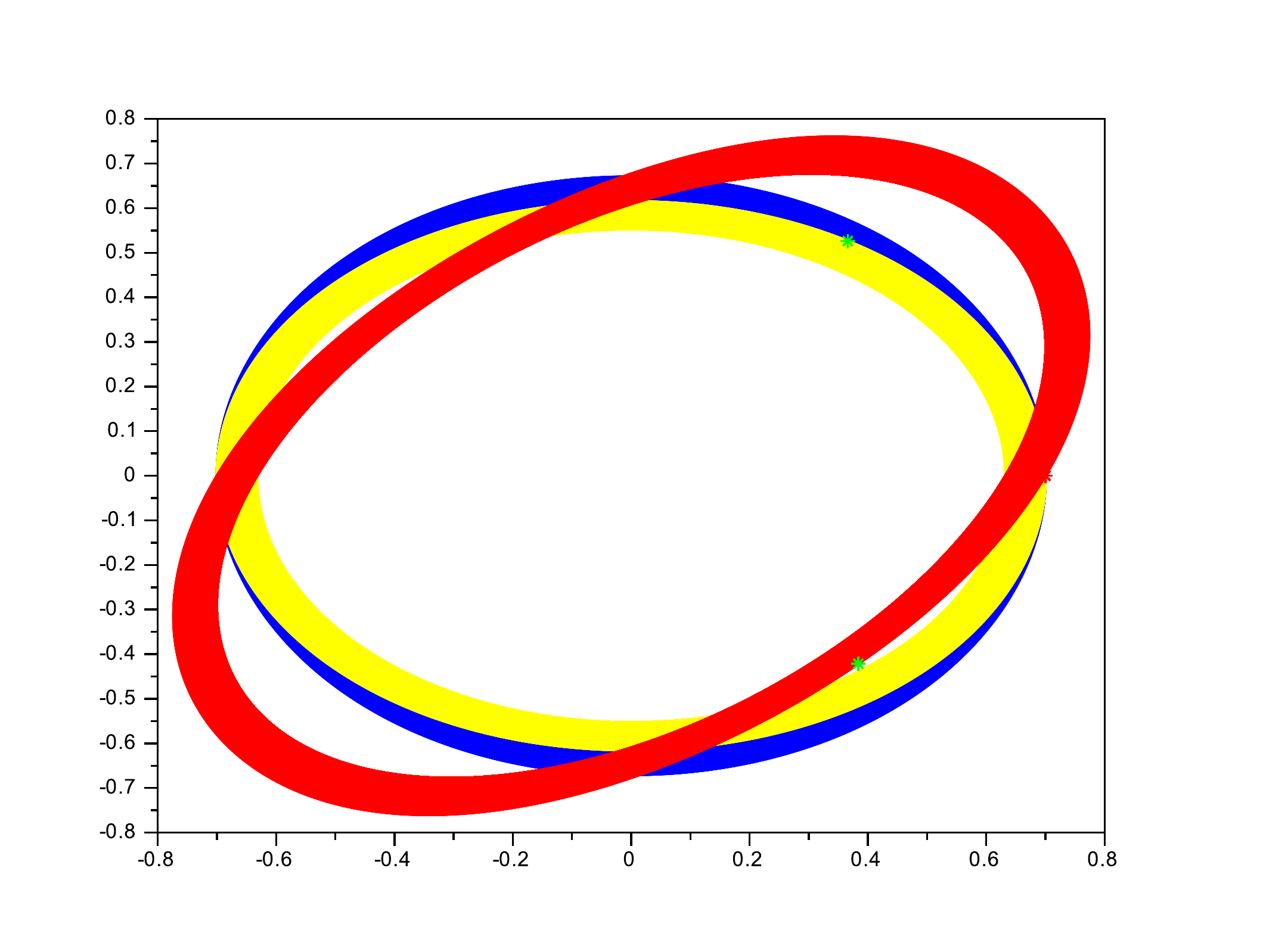}
                \caption{Symplectic Euler (red), Verlet (yellow), Forest-Ruth (blue), $\Delta t = 0.9$ }
                \label{fig:se_v_fr_CC}
        \end{subfigure}
        \caption{Higher order integrators (phase space diagrams) }\label{fig:higher_order_CC}
\end{figure}

As an additional benchmark study, we have used the pendulum equation, $\ddot{\theta} = -\sin (\theta)$. The explicit Euler is clearly the most unstable, introducing ghost energies as seen in the phase space diagram comparison with respect to the midpoint method in figure \ref{fig:ee_mp_CC}. On the other hand, in figure \ref{fig:sm_se_CC}, the symplectic Euler and Midpoint methods have a much higher stability range, with the latter yielding slightly lower energy variations. As second order explicit integrators, the modified half step method is also more stable than the original version (as depicted in figure \ref{fig:hs_mhs_CC}). Nevertheless, both methods tend to add ghost energies,  but are much more stable than the first order explicit Euler. As the order of accuracy increases, larger time steps can be used (as it is the case with Runge-Kutta methods), but the performance impact does not justify such a trade-off. Finally, for models where no damping or non-conservative forces are involved, the Forest-Ruth and 
Verlet methods are the natural choices (comparative phase space plots in figure \ref{fig:se_v_fr_CC}). Nevertheless, in case accuracy can be sacrificed, we still recommend using the first order symplectic Euler method as it has the highest stability versus complexity score.

\section{Conclusion}
\label{sec:conclusion_CC}
In this work we have discussed the implications of employing an explicit integration method for updating a soft or semi-rigid body simulation. For real time applications where accuracy is not a goal, we recommend using a symplectic integrator as it has the best performance and stability score. Even when non-conservative forces are involved, this class of integrators is able to cope with stiff constraints. However, for applications where accuracy cannot be sacrificed, both the Runge-Kutta or the Forest-Ruth integrators can be used. The latter choice is accurate up to the third order and conserves energy, while the Runge-Kutta progressively loses small amounts of energy, counting for a slight increase in stability.

On a final note, implicit integrators, mentioned in the related work section \ref{sec:related_work_CC}, are not the usual choice for real time applications, thus motivating our investigation towards an explicit integrator alternative.

\begin{appendices}
\section{Explicit Integration Methods}
  In the following appendix, we present the minor modifications of the explicit integration methods that were tested in our simulation application. These numerical methods make use of an acceleration function,
  $f$, a fixed time-step, $h$, and compute new positions and velocities, given the previous state $(\mathbf{x}_n, \mathbf{v}_n)$. 
\begin{table}[h]
\centering 
\begin{tabular}{| >{$}l<{$} | >{$}l<{$} | >{$}l<{$}| }
\hline
\text{Symplectic Midpoint} & \text{Modified Half Step}\\
\hline
\mathbf{v}_{n+1} = \mathbf{v}_n + \frac{h}{2} f(\mathbf{x}_n, \mathbf{v}_n) &
\mathbf{v}_{n+\frac{1}{2}} = \mathbf{v}_n + \frac{h}{2} f(\mathbf{x}_n, \mathbf{v}_n) \\

\mathbf{x}_{n+1} = \mathbf{x}_n + h \mathbf{v}_{n+1} &
\mathbf{x}_{n+\frac{1}{2}} = \mathbf{x}_n + \frac{h}{2} \mathbf{v}_{n+\frac{1}{2}}\\
& \\
&
\mathbf{v}_{n+1} = \mathbf{v}_n + hf(\mathbf{x}_{n+\frac{1}{2}},\mathbf{v}_{n+\frac{1}{2}}) \\
& 
\mathbf{x}_{n+1} = \mathbf{x}_n + h \mathbf{v}_{n+\frac{1}{2}}\\
\hline
\end{tabular}
\caption{Slightly modified integrators for improved stability and accuracy} 
\label{tab:appendix_CC} 
\end{table}

\begin{figure}[h]
        \centering
        \begin{subfigure}[b]{0.5\textwidth}
                \centering
                \includegraphics[width=\textwidth]{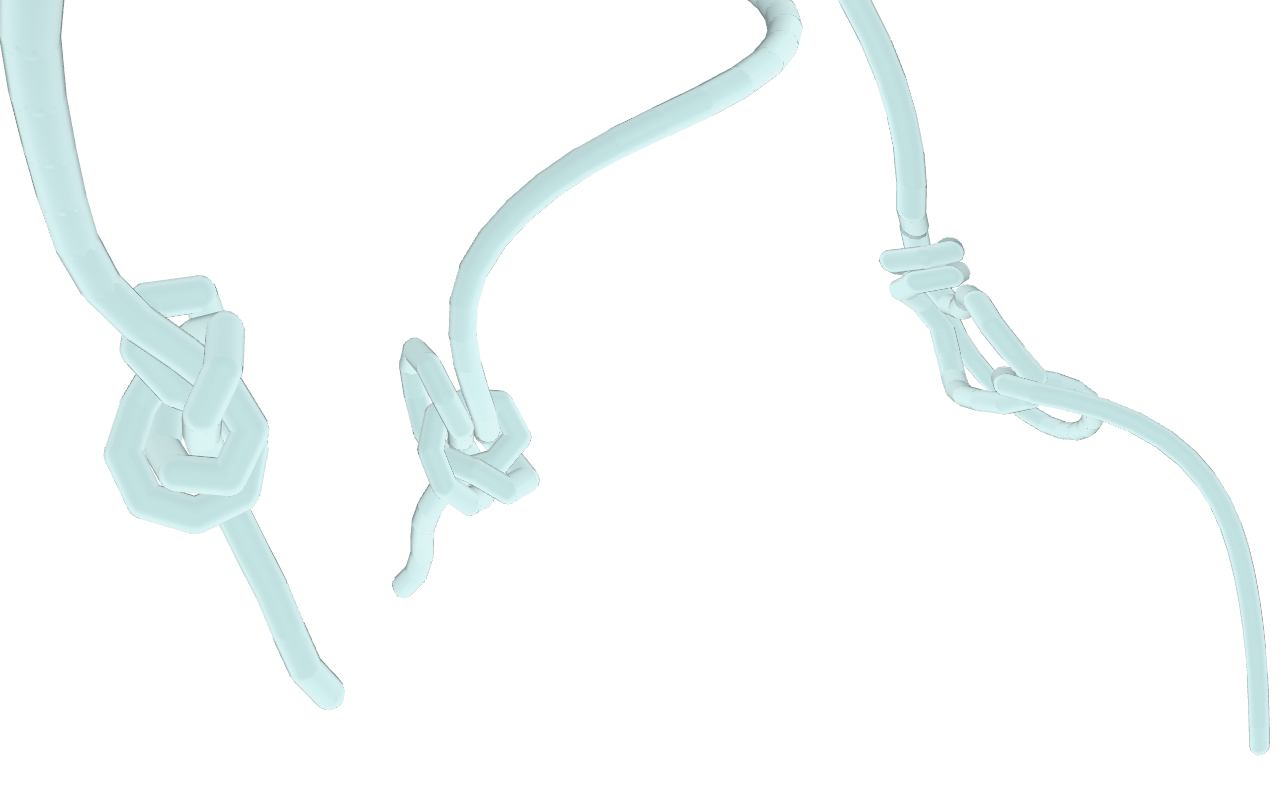}
                \caption{Creating complex knots: tight knots tend correspond to very high elastic potentials. The numerical stability is crucial for plausible knot behaviour. }
                \label{fig:knotting_CC}
        \end{subfigure}%
        ~ 
        \begin{subfigure}[b]{0.5\textwidth}
                \centering
                \includegraphics[width=\textwidth]{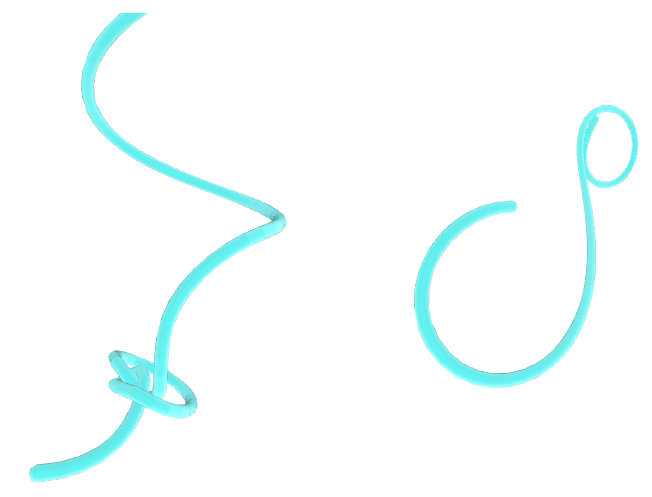}
                \caption{Highly elastic cable unknotting itself under the action of strong constraint forces}
                \label{fig:unknotting_CC}
        \end{subfigure}
        \caption{Hose object knot tying}\label{fig:knotting_unknotting_CC}
\end{figure}

\end{appendices}
\newcommand{\etalchar}[1]{$^{#1}$}

\end{document}